\documentclass[runningheads]{llncs}
\usepackage{graphicx}
\usepackage{enumitem}
\usepackage{booktabs}
\usepackage{adjustbox}

\usepackage{hyperref} 
\hypersetup{hidelinks} 

\begin{document}
%


\title{Evaluating Microservice Organizational Coupling based on Cross-service Contribution}

%
%
\author{Xiaozhou Li\inst{1} \and
Dario Amoroso d’Aragona\inst{2} \and
Davide Taibi\inst{1,2}}
\authorrunning{F. Author et al.}
%
\institute{University of Oulu, Oulu, Finland \\
\email{firstname.lastname@oulu.fi}\\
\and
Tampere University, Tampere, Finland \\
\email{firstname.lastname@tuni.fi}\\
}
\maketitle              
\begin{abstract}

For traditional modular software systems, "high cohesion, low coupling" is a recommended setting while it remains so for microservice architectures. However, coupling phenomena commonly exist therein which are caused by cross-service calls and dependencies. In addition, it is  noticeable that teams for microservice projects can also suffer from high coupling issues in terms of their cross-service contribution, which can inevitably result in technical debt and high managerial costs. Such organizational coupling needs to be detected and mitigated in time to prevent future losses. Therefore, this paper proposes an automatable approach to evaluate the organizational couple by investigating the microservice ownership and cross-service contribution. 

\end{abstract}
\section{Introduction}

Together with the advance of software engineering theories and practice, modularization has long been considered a mechanism to enhance a system's flexibility and comprehensibility 
system as well as its development efficiency \cite{parnas1972criteria}. Meanwhile, coupling and cohesion are the two critical concepts for modularized systems that characterize the interdependence amongst the modules when one well-recognized software design principle is ``high cohesion low coupling". Especially, microservice, as one of the most dominantly popular modularized architectures for cloud-native systems, is also required to comply with the principle in order to guarantee the architecture quality \cite{walpita2020}. Coupling is a common issue for microservice systems when many recent studies have proposed definitions as well as methods to identify and evaluate different types of coupling therein \cite{d2023microservice,zhong2023measuring}. 

Despite the importance of the issue, limited studies have been conducted on handling the couplings in microservice architecture. For example, Zhong et al. propose the Microservice Coupling Index (MCI) based on relative measurement theory which measures the dependence of the target microservices relative to the possible couplings between them \cite{zhong2023measuring}.  d’Aragona et al. propose to use commit data as a metric to statically calculate logical coupling between microservices and validate the existence of such couplings in a large number of open-source microservices projects \cite{d2023microservice}. Specially, these studies propose microservice couplings from dynamic analysis or static analysis perspectives, as well as the temporal and deployment perspectives \cite{walpita2020}. However, limited studies have taken into account the couplings on the organizational level, though organization-related issues are usually as important as technology issues if not more so \cite{perry1994people}. 

For large software projects, properly structured organization shall contribute to effective collaboration with reduced communication, which is critical for the project’s success \cite{brooks1995mythical}. For microservice-based projects, stakeholders shall be aware of and able to handle critical organizational issues, e.g., coupling, for the migration from monolith to microservices \cite{newman2021building}. As microservice promotes and benefits from ``strong module boundaries", the communication structure of the organization building it shall mirror such structure with the boundaries \cite{conway1968committees,Fowler201507}. After all, a module, in many contexts, is considered more than just a subprogram but rather a responsibility assignment \cite{parnas1972criteria}. It implies that the organizational structure of microservice projects shall also establish boundaries amongst different teams where developers within a team shall closely collaborate (i.e., high cohesion) while developers across teams shall be highly independent (i.e., low coupling). Therefore, it is not surprising that the notion of ``One Microservice per Developer" has been promoted by many practitioners and companies \cite{balalaie2016microservices,richardson2022dark,Reinfurt2021,daragona2023one}. Though several studies have investigated microservice projects' organizational structure \cite{bavskarada2018architecting,li2023analyzing}, studies on the coupling of microservices in terms of their organization structures are still limited.

Therefore, in this study, we propose the metric to assess the coupling on organizational structure level for microservice projects, named \textit{organizational coupling}. A prerequisite step of evaluating such coupling is to identify the team for each microservice of the target project. Therefore, the degree to which two microservices are coupled in terms of developers' ``cross-boundaries" contribution can be determined by that of those developers simultaneously belonging to both teams. To such an end, our work here can answer the following research question (RQ): \textit{How to evaluate the organizational coupling between microservices in terms of cross-service contribution?} 


The remainder of this paper is organized as follows. Section 2 introduces the related studies regarding coupling in microservice and microservice organizational structure. Section 3 presents the method to evaluate the organizational coupling between microservices. Section 4 uses a case study to validate the method in terms of its operationality. Section 5 provides a discussion on the implication, limitations, and future work when Section 6 concludes the paper.

\section{Related Work}

The organizational structure of software projects has long been a critical factor determining the projects' success \cite{brooks1995mythical}. Many studies have proposed approaches to analyze or improve software projects' organizational structure. Nagappan et al. propose a metric scheme to quantify organizational complexity regarding the product development process checking if the metrics impact failure-proneness where the level of organizational code ownership is a key metric\cite{nagappan2008influence}. Mockus studies the relationship between developer-centric measures of organizational change and the probability of customer-reported defects with the results showing organizational change is associated with lower
software quality \cite{mockus2010organizational}. Isern et al. investigate the popular agent-oriented methodologies in terms of their support and possibilities for modeling organizational structures with different levels of complexity \cite{isern2011organizational}. 

Regarding the organizational structure of microservice projects, Li et al. propose an approach using social network analysis (SNA) to reconstruct the organizational structure of microservice-based software projects in terms of contributor collaboration \cite{li2023analyzing}. d'Aragona et al. investigate the application of the ``one microservice per developer" principle in OSS microservice projects and propose an approach of using exploratory factor analysis (EFA) to establish the different team specialty profiles \cite{daragona2023one}. Ashraf et al. conduct an empirical study and find that developer communities change considerably through projects' lifetime and that their alignment with the pre-defined microservice (or subsystem) teams is mostly low \cite{ashraf2021communities}. 

On the other hand, many studies have proposed methods to measure the coupling between software modules. Allen et al. propose related information theory-based measures of coupling and cohesion of a module based on the properties proposed by Briand et al. \cite{briand1996property,allen2001measuring}. Poshyvanyk and Marcus also propose a new set of coupling measures for object-oriented systems, named conceptual coupling, based on the semantic 
information shared between elements of the source 
code \cite{poshyvanyk2006conceptual}. Other methods are also proposed to measure the coupling between packages or classes \cite{gupta2009package,hammad2014framework}. All such coupling metrics and proposed measuring methods focus on the dependency relations within the source code without considering the connections among developers or latent teams. 

Regarding the coupling in microservice-based systems, Zhong et al. propose Microservice Coupling Index (MCI) derived from the relative measurement theory, which measures how the coupled microservices are relative to the possible couplings between them \cite{zhong2023measuring}. Pedraza-Coello and Valdés-Souto propose a method to measure the coupling between microservices in early phases based on COSMIC method concepts regarding the data movements in functional processes \cite{pedraza2021measuring}. d'Aragona et al. propose a metric to statically calculate logical coupling between microservices based on commits \cite{d2023microservice}. Though these studies have addressed the issue of coupling in microservice-based systems, limited have yet considered the couplings on the organizational level. 

\section{Organizational Coupling}

Here we introduce the concept of organizational coupling and the methodology to evaluate the organizational coupling in any particular microservice-based system, which answers the proposed research question. 



\subsection{Identify Microservice Teams}

As the initial step of evaluating the coupling between microservice teams, it is necessary to have a method to identify the team for each microservice in the target project. To do so, we adapt the method proposed by Bird et al. regarding the ownership profile of a particular software component~\cite{bird2011don}, which, herein, is the microservices. 

Let $M$ be the target microservice of a particular software product where file set $F$ is identified in the folder (or the repository) of $M$ located in the project. Thus, we see all the contributors that have committed to any of those $n$ files establishing the team of microservice $M$, denoted as $T_M$. 

\begin{figure}[!ht]
    \centering
    \includegraphics[width=\linewidth]{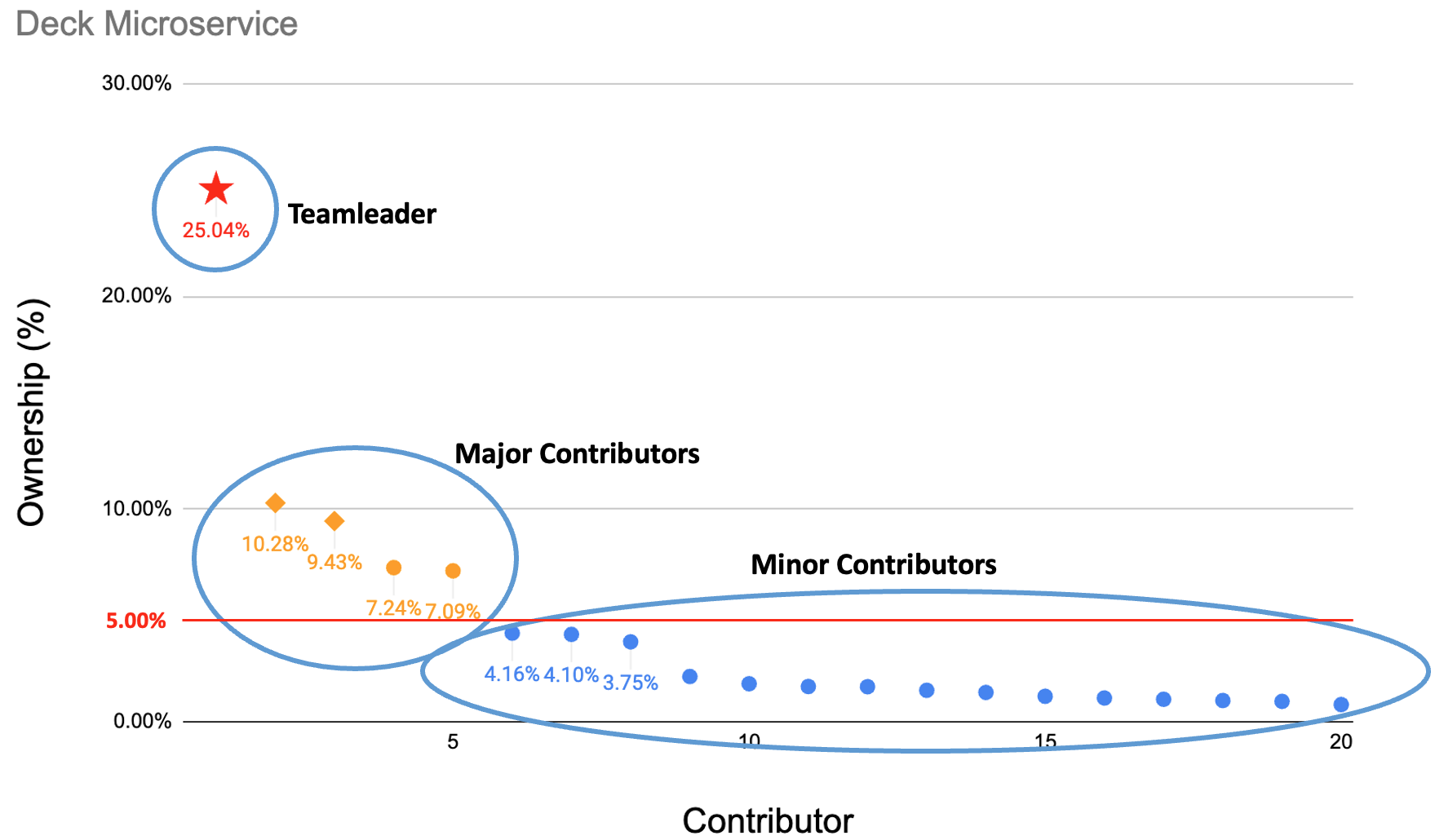}
    \caption{Ownership Proportion Example}
    \label{fig:spnkrdeckowner}
\end{figure}

Herein, we calculate the quantified contribution of any developer $D \in T_M$ to $M$ as the sum of all the number of changes to each file $f \in F$. Furthermore, we calculate $D$'s proportion of ownership (or ownership) of $M$ as the ratio of the number of commit changes that $D$ has made relative to the total number of commit changes (in terms of lines of code) for $M$. To be noted, compared to the original study of Bird et al. \cite{bird2011don}, we use the number of commit changes instead of the number of commits due to the consideration that commits vary largely between one and another in terms of the exerted effort from the developers. 

Therefore, based on the calculated ownership proportion of each $D \in T_M$, we see the developer(s) who has the highest proportion of ownership for $M$ as the \textit{Teamleader(s)}. According to the definitions by Bird et al. \cite{bird2011don}, we also define 1) \textit{Major} contributors as the developers whose contribution reach at least 5\% proportion level, and 2) \textit{Minor} contributors as the developers whose contribution reach at least 5\% proportion level. An example of microservice team with ownership proportion is shown in Figure \ref{fig:spnkrdeckowner}. In this way, for each microservice in a given microservice-based architecture, based on the commits data, we can identify the team, i.e., all the developers who have contributed to it, and the ownership proportion of each developer, i.e., how much contribution ratio his/her is to the whole team. 

\subsection{Contribution Switch as Weight}

Herein, we also take into account the phenomenon of the developer's contribution switch as an important factor influencing the organizational coupling between microservices. On the organizational level, we consider two individual microservices (as well as their teams) are more heavily coupled when the developers from either team more frequently commit to the other.

\begin{figure}[!ht]
    \centering
    \includegraphics[width=\linewidth]{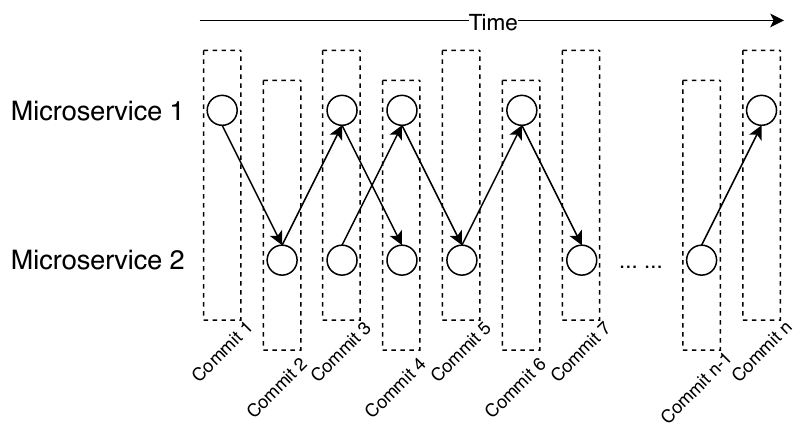}
    \caption{Contribution Switch between Microservices}
    \label{fig:switch}
\end{figure}

Given two microservices $M_a$ and $M_b$, assume a developer $D \in T_{M_a}$ or $D \in T_{M_b}$ whose \textit{contribution switch weight} between these two microservices is denoted as $S_{D}(M_{a}, M_{b})$. Therefore, whenever $D$ commits to $M_a$ and then commits to $M_b$ afterward (e.g., Commit 1 and 2 in Figure \ref{fig:switch}), we consider such an incidence as a \textit{contribution switch} of developer $D$ from $M_a$ to $M_b$. Similarly, developer $D$ also switches from $M_b$ back to $M_a$ via Commit 3 shown in Figure \ref{fig:switch}. To be noted, herein we only take into account the sequential relation of the commit series without considering the time intervals in between. 

Therefore, given the sequence of commits of $D$ in terms of $M_a$ and $M_b$, 
we can simply count the number of contribution switches therein. In addition, regarding the situation of logically coupled commits \cite{d2023microservice} where both microservices are changed in a single commit (e.g., Commit 3 and 4), we consider this situation as two contribution switches. 

To generalize, given the previously described situation where $k$ contribution switches are performed by $D$ between $M_a$ and $M_b$ while $D$ has in total $n$ commits for both microservices, the contribution switch weight can be calculated as follows. 

\begin{equation}
    S_{D}(M_{a}, M_{b}) = \frac{k}{2\times{(n-1)}}
\end{equation}

Taking Fig. \ref{fig:switch} as an example where $n = 8$, as we can observe eight contriution switches (i.e., $k = 8$), $S_{D}(M_{1}, M_{2}) = 8/(2\times{(8-1)}) = 0.571$. Considering the situation where every commit from the developer changes both microservices (i.e., logical coupling \cite{d2023microservice}), $S_{D}(M_{1}, M_{2}) = 14/(2\times{(8-1)}) = 1$. On the contrary, when the developer only contributes to one microservice, $S_{D}(M_{1}, M_{2}) = 0/(2\times{(8-1)}) = 0$. It means the two microservices are not coupled in terms of the contribution of $D$ on the organizational level. Therefore, we can easily conclude that $S_{D}(M_{a}, M_{b}) \in [0, 1]$. To be noted, the contribution switch weight is only to influence the organizational coupling in terms of individual developers.

\subsection{Measure Organizational Coupling}

Given any two microservices $M_a$ and $M_b$, $T_{M_a}$ and $T_{M_b}$ are the teams for each microservice respectively, which are identified by the method proposed in Section 3.1. Therein, we can simply identify the $p$ developers who have contributed in both microservices, denoted as $T_{(M_a \cap M_b)} = \{D_{1}, D_{2}, ... D_{p}\}$. For any particular developer $D_i \in T_{(M_a \cap M_b)}$, all the commits he/she has conducted in temperal sequence are denoted as $C_{D_i}$. For each $c \in C_{D_i}$, we can identify on which microservice it commits to. Therefore, by finding the ones that are committed to $M_a$ or $M_b$ or both, we obtain a sub-sequence of commits, denoted as $C_{D_i}(M_{a}, M_{b})$. Such a commit sequence can be depicted as a figure similar to Fig. \ref{fig:switch} where all the contribution switches can be identified with the contribution switch weight, $S_{D_i}(M_{a}, M_{b})$, calculated based on the method described in Section 3.2.


To investigate the coupled contribution of $D_i$ on $M_{a}$ and $M_{b}$, we adopt the harmonic mean of $D_i$'s contribution in them, considering the reason that the more equally any developer commits to multiple microservices, the more organizationally coupled the two microservices are, regarding this developer's contribution. 

Let $\{ca_1, ca_2, ... ca_m\}$ be the corresponding contribution value sequence for the $m$ commits in $C_{D_i}(M_{a})$ while $\{cb_1, cb_2, ... cb_n\}$ be that for the $n$ commits in $C_{D_i}(M_{b})$. Herein, the contribution value of each commit is calculated by the sum of all the number of changes to each file in the target microservices. Let $OC(D_{i}, M_{a}, M_{b})$ be the organizational coupling (OC) caused by developer $D_i$'s cross-service contribution on microservices $M_{a}$ and $M_{b}$, we can calculate $OC(D_{i}, M_{a}, M_{b})$ as follows.

\begin{equation}
    OC(D_{i}, M_{a}, M_{b}) = (\frac{2\sum_{j=1}^{m}ca_j\sum_{k=1}^{n}cb_j}{\sum_{j=1}^{m}ca_j + \sum_{k=1}^{n}cb_j}) \times S_{D_i}(M_{a}, M_{b})
\end{equation}

Thus, the overall organizational coupling between $M_{a}$ and $M_{b}$, denoted as $OC(M_{a}, M_{b})$, can be calculated as follows.

\begin{equation}
    OC(M_{a}, M_{b}) = \sum_{i=1}^{p}OC(D_{i}, M_{a}, M_{b}) = \sum_{i=1}^{p}(\frac{2\sum_{j=1}^{m}ca_j\sum_{k=1}^{n}cb_j}{\sum_{j=1}^{m}ca_j + \sum_{k=1}^{n}cb_j}) \times S_{D_i}(M_{a}, M_{b})
\end{equation}

\section{Case Study}

In this study, we demonstrate the applicability of the proposed organizational coupling evaluation method with a case study. We select, \textit{Spinnaker}\footnote{\url{https://spinnaker.io/}}, a microservice-based application management and deployment system supporting software change releases. Spinnaker is an open-source, multi-cloud continuous delivery platform that combines a powerful and flexible pipeline management system with integrations to the major cloud providers. Herein, we use the Spinnaker project as a proof-of-concept to demonstrate and validate how to identify and evaluate the organizational coupling within microservice-based systems.

\subsection{Data Collection}

\begin{figure}[!ht]
    \centering
    \includegraphics[width=\linewidth]{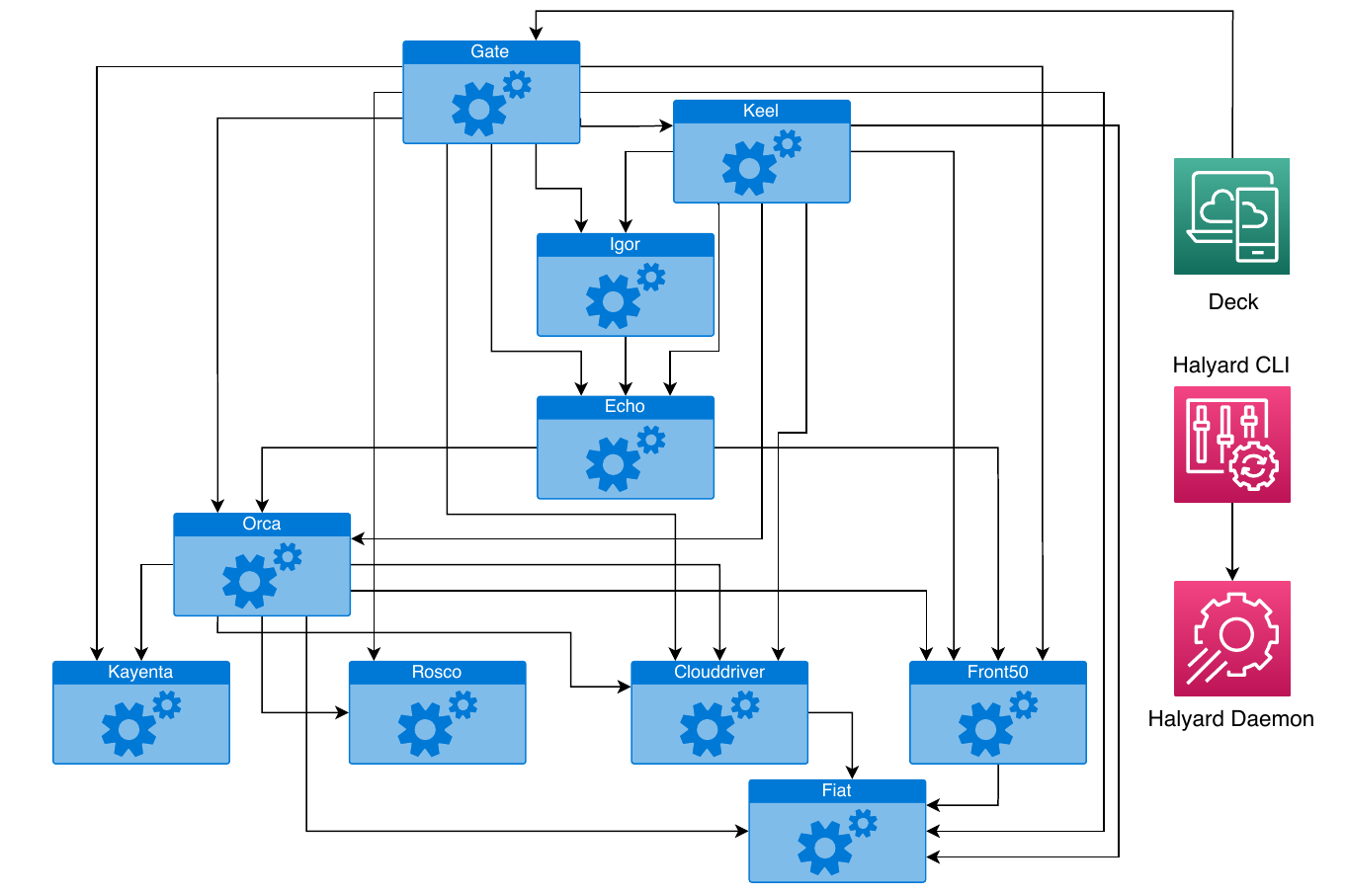}
    \caption{Spinnaker Architecture Overview}
    \label{fig:spinnakerms}
\end{figure}

Spinnaker contains 12 independent microservices\footnote{\url{https://spinnaker.io/docs/reference/architecture/microservices-overview/}}. The dependencies of the microservices are shown in Figure \ref{fig:spinnakerms}. 

The 12 microservices include \textit{CloudDriver}, \textit{Deck}, \textit{Echo}, \textit{Fiat}, \textit{Front50}, \textit{Gate}, \textit{Halyard}, \textit{Igor}, \textit{Kayenta}, \textit{Keel}, \textit{Orca}, and \textit{Rosco}. The detailed functionality and responsibility of each microservices are introduced in the Spinnaker official documentation as well as its GitHub repositories\footnote{\url{https://github.com/spinnaker}}. To be noted, different from other popular microservice-based projects, e.g., eShopOnContainers\footnote{\url{http://github.com/dotnet- architecture/eShopOnContainers}}, Spinnaker project is organized as polyrepo architecture instead of monorepo \cite{brousse2019issue}. Therefore, we shall gather data from the 12 corresponing repositories of the project. 

By using the GitHub REST API \footnote{\url{https://docs.github.com/en/rest?apiVersion=2022-11-28}}, we are able to collect all the commit data for the target 12 microservices of Spinnaker project. We collect the 43,654 commits from all 12 microservice repositories between 2012-03-18 and 2023-07-06. 801 different developers contributed to all these commits with 241,828 file changes. 


\begin{figure}[!ht]
    \centering
    \includegraphics[width=0.8\linewidth]{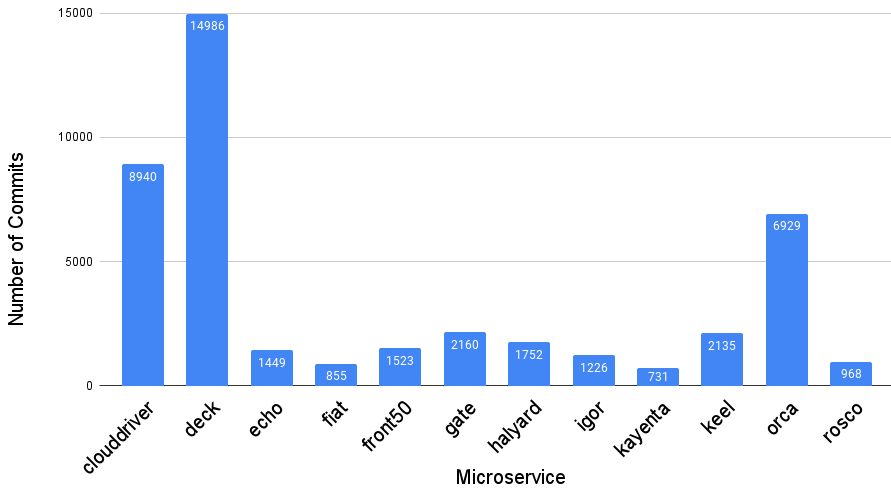}
    \caption{Number of Commits for each Microservice}
    \label{fig:compms}
\end{figure}

\begin{figure}[!ht]
    \centering
    \includegraphics[width=0.8\linewidth]{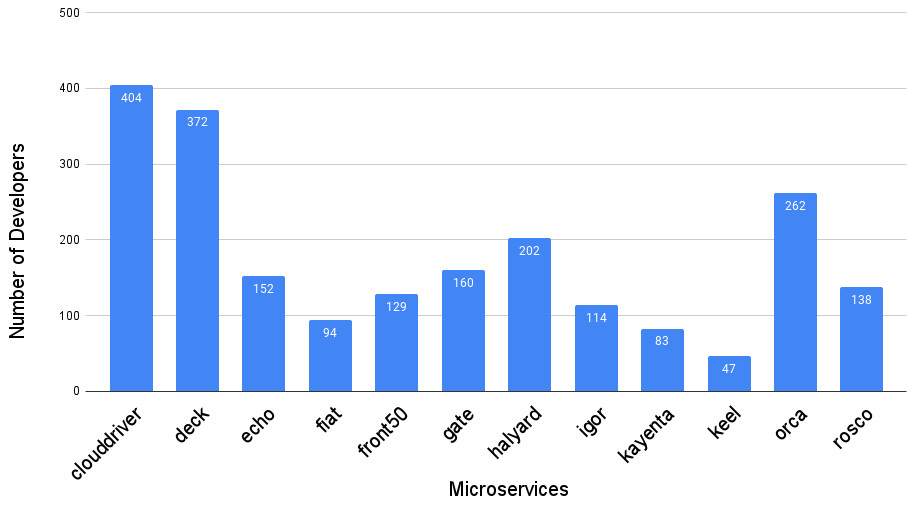}
    \caption{Number of Developers for each Microservice}
    \label{fig:devpms}
\end{figure}

The distribution of 1) the number of commits for each microservice and 2) the number of different developers for each microservice are shown in Figure \ref{fig:compms} and Figure \ref{fig:devpms}. To be noted, in the original dataset, for each individual commit the contributor is identifed by \textit{author\_email}. However, considering the situation where multiple emails can belong to the same user, e.g., \textit{lwander@users.noreply.github.com} and \textit{lwander@google.com}, we preprocess the author identity by dropping the email extention and combining such accounts. 


\subsection{Results}

\subsubsection{Identify Microservice Teams}

Firstly we identify the developer team for each microservice using the method introduced in Section 3.1. Due to the fact that Spinnaker project is structured as poly-repo, each microservice is an independent repository. Therefore, the team of each microservice shall contain all the contributors of each repository, which is comparatively easier to identify compared to mono-repo projects, e.g., eShopOnContainer. The number of developers in each microservice team is shown in Figure \ref{fig:devpms}.

\begin{figure}[!ht]
    \centering
    \includegraphics[width=0.9\linewidth]{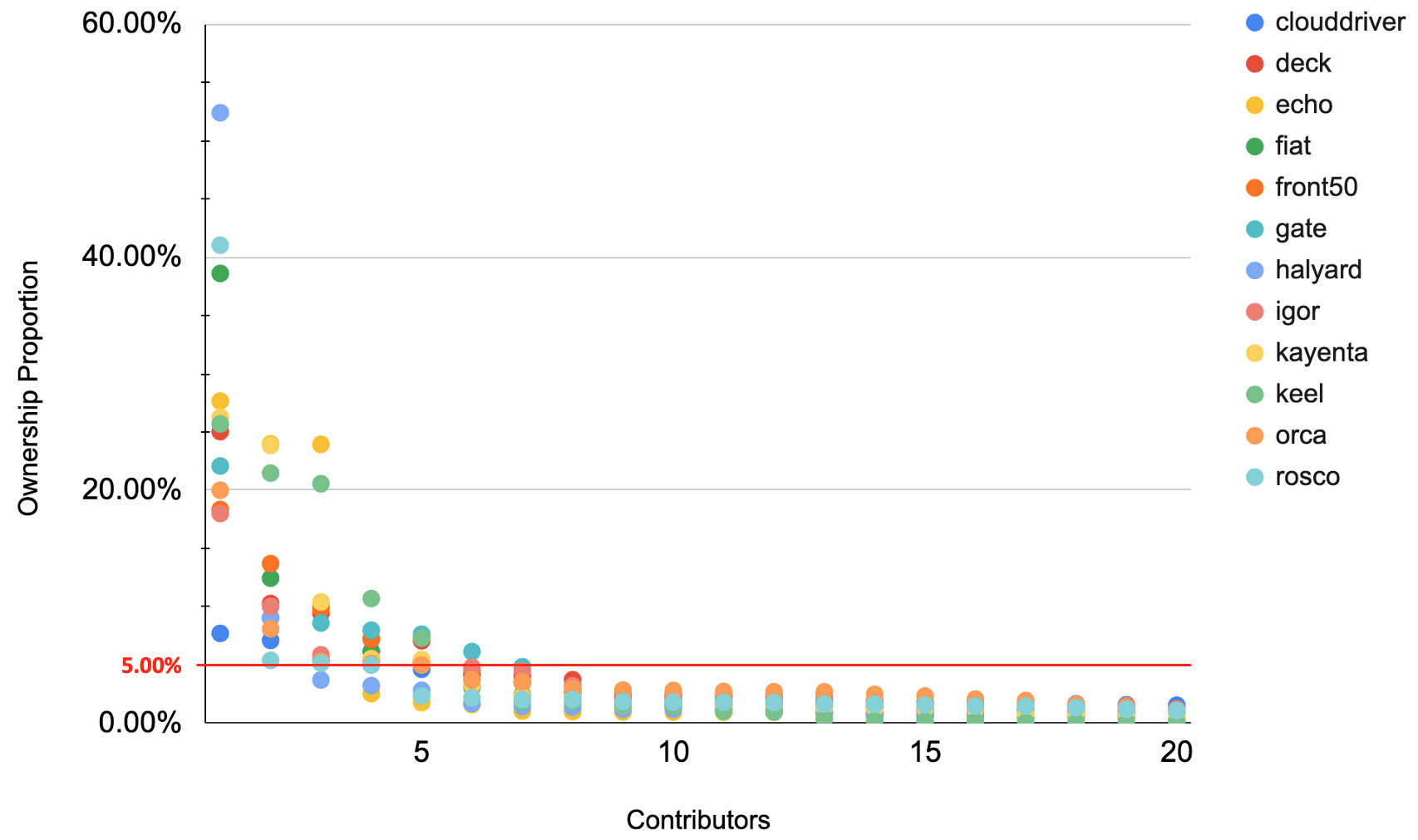}
    \caption{Ownership Proportion of Spinnaker Microservices}
    \label{fig:spnkrowners}
\end{figure}

In addition, we can also further specify the team structure of each microservice team by identifying the teamleaders, major contributors and minor contributors of each team. Figure \ref{fig:spnkrowners} shows the Top 20 contributors of each microservice team in terms of their ownership proportion. It is easy to observe that all microservice teams have at least one teamleader and one major contributor. Meanwhile, no team has more than six major contributors (including the Teamleader). 

\begin{table}[!ht]
    \centering
    \adjustbox{max width=\textwidth}{
    \begin{tabular}{|l|l|l|}
        \toprule
        \textbf{Microservice} & \textbf{Teamleader} & \textbf{Major Contributor(s)}\\
        \midrule
        CloudDriver & cfiexxx (7.72\%) & duftxxx (7.12\%), lwanxxx	(5.47\%), camexxx (5.18\%) \\
        Deck&chrixxx (25.04\%)&vmurxxx (10.28\%), benjxxx (9.43\%), githxxx (7.24\%), zantxxx (7.09\%)\\
        Echo&clinxxx (27.65\%)&ajorxxx (23.98\%), adamxxx (23.94\%)\\
        Fiat&ttomxxx (38.60\%)&devexxx (12.46\%), cfiexxx (10.15\%), rziexxx (6.19\%), adamxxx (5.39\%)\\
        Front50&adamxxx (18.35\%)&rziexxx	(13.70\%), ajorxxx (9.82\%), danixxx (7.22\%), cfiexxx (5.12\%)\\
        Gate&danixxx (22.08\%)&adamxxx	(9.04\%), jacoxxx (8.61\%), builxxx	(7.99\%), ajorxxx (7.65\%), cfiexxx (6.16\%)\\
        Halyard&lwanxxx (52.39\%)&ezimxxx	(9.06\%)\\
        Igor&rziexxx (18.00\%)&jorgxxx (10.03\%), ezimxxx (5.88\%), clinxxx (5.56\%), tomaxxx (5.31\%)\\
        Kayenta&duftxxx	(26.25\%)&fielxxx	(23.84\%), explxxx (10.42\%), asmixxx (5.55\%), chrixxx (5.48\%)\\
        Keel&1323xxx (25.69\%)&lhocxxx	(21.47\%), rflexxx (20.56\%), robfxxx	(10.71\%), emjexxx (7.31\%)\\
        Orca&rflexxx (20.00\%)&robfxxx	(8.11\%), clinxxx (5.32\%), robxxx (5.14\%), adamxxx (5.01\%)\\
        Rosco&duftxxx (41.03\%)&ezimxxx (5.41\%), ttomxxx (5.16\%), andexxx	(5.00\%)\\
        \bottomrule
    \end{tabular}
    }
    \caption{Microservice Teamleaders and Major Contributors}
    \label{tab:msteams}
\end{table}
\vspace{-0.7cm}

Specifically, we list the teamleader and major contributors of each microservice team in Table \ref{tab:msteams}. Considering the privacy reason, we only show the first four letters of each contributor's identity. We can observe that majority of the teamleaders have 20\% - 30\% ownership proportion. The teamleader of Halyard microservice has the highest ownership of the service (52.39\%) when the teamleader of CloudDriver microservice has the lowest (7.72\%). Meanwhile, we can also observe that majority of the microservice teams have one clear teamleader whose ownership proportion is at least 5\% higher than that of the second major contributor. Five microservice teams have at least two contributors who share similar ownership proportion. 

Moreover, it is also noticeable that developer \textit{duftxxx} is the teamleader of both \textit{Kayenta} service and \textit{Rosco} service. He/she is also the major contributer of \textit{CloudDriver} service. Meanwhile, 11 out of the 12 microservice teamleaders are also major contributor of at least one other team. 

\subsubsection{Organizational Couplings and Evolution}

\begin{figure}[!ht]
\centering
  \includegraphics[width=\textwidth]{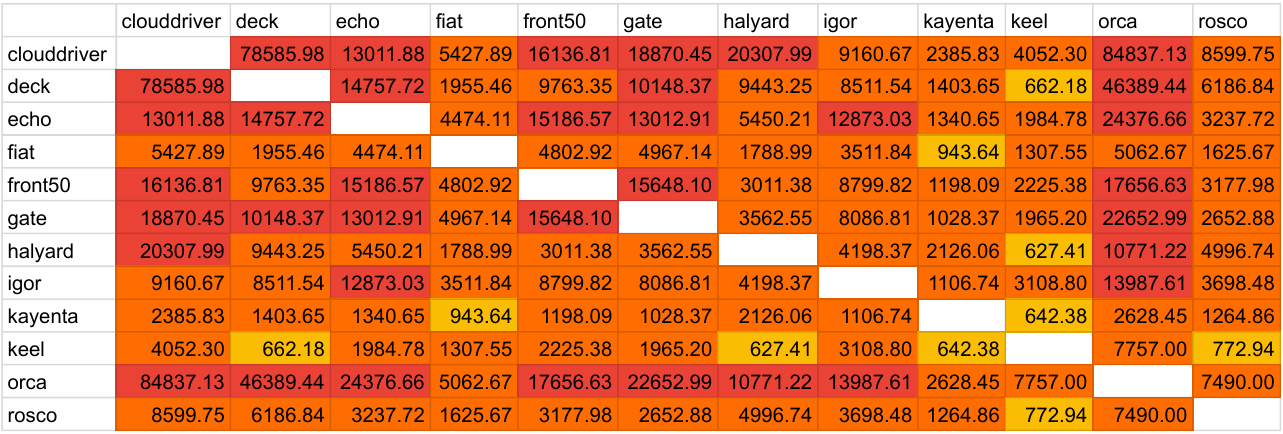}
  \caption{Organizational Coupling between Services (Version 1.30.2)}
  \label{fig:heatmap}
\end{figure}

With the team of each microservice identified, we can then calculate the organizational coupling between each pair of them by adopting the method introduced in Section 3.3. According to the version log of Spinnaker\footnote{\url{https://spinnaker.io/docs/releases/versions/}}, the latest stable version (1.30.2) was released on June 1st, 2023. We select all the commit data until this date and calculate all the organizational couplings. We set the coloring criteria as: 1) Red (Very Highly Coupled) : $OC \geq 10,000$; 2) Orange (Highly Coupled): $1,000 \leq OC < 10,000$; 3) Yellow (Loosely Coupled): $100 \leq OC < 1,000$; 4) Green (Very Loosely Coupled): $OC < 100$. The results are shown in Figure \ref{fig:heatmap}.

We can observe that majority of the 12 microservices of Spinnaker are at least highly coupled in terms of developers' cross-service contribution. The most severely high coupling is the one between \textit{Orca} service and \textit{CloudDriver} service (84837.13). Meanwhile, both these two services are also heavily coupled with all other services. On the contrary, \textit{Keel} is loosely coupled with several services, including \textit{Deck}, \textit{Halyard}, \textit{Kayenta} and \textit{Rosco} when \textit{Kayenta} and \textit{Fiat} are also loosely coupled. 

\begin{figure}[!ht]
\centering
  \includegraphics[width=\textwidth]{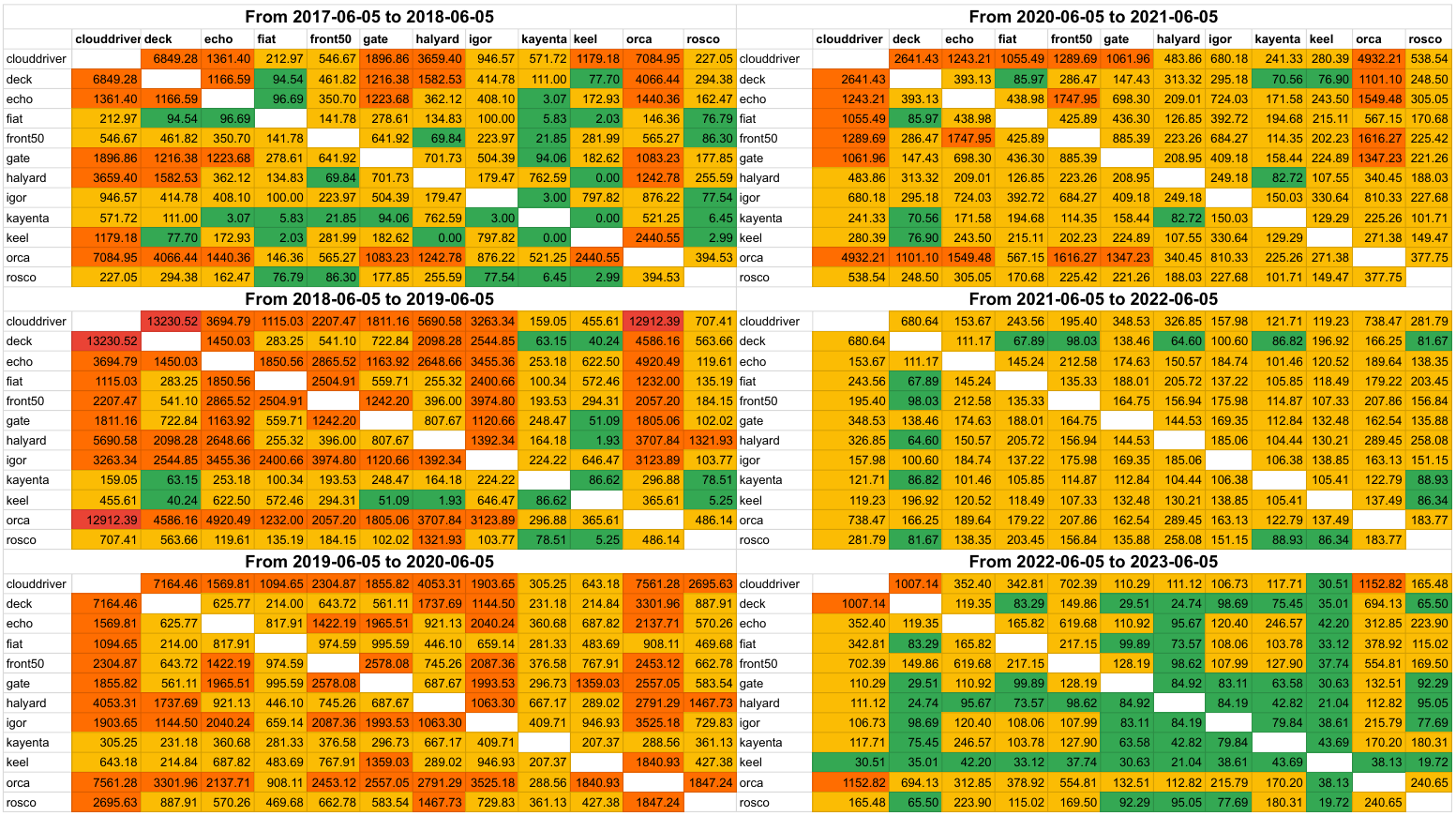}
  \caption{Evolution of Organizational Coupling between Services}
  \label{fig:heatmapevo}
\end{figure}

Such a phenomenon likely results from the fact that the Spinnaker project had the first initial release in late 2015 with the earliest service repository created in May 2014. It is only reasonable that in the early development phase, a limited number of developers were heavily involved in nearly all the services. Therefore, we can also investigate the changes in such organizational coupling between services through the project timeline. The first stable version (Version 1.0.0) of Spinnaker was released on June 5th, 2017. We select six commit datasets of six consecutive years from 2017-06-05 to 2023-06-05. By adopting the same method for each dataset, we can obtain six different heatmaps regarding the organizational coupling between services in the specific one-year period and observe the changes (shown in Figure \ref{fig:heatmapevo}).

Observing the service organizational coupling from 2017-06-05 to 2018-06-05, we find that \textit{Kayenta} and \textit{Keel} are very loosely coupled with the majority of the others. The reason is likely \textit{Kayenta} was created in January 2017 while \textit{Keel} in October 2017. \textit{CloudDriver} and \textit{Orca} are still highly coupled with many other services. Thereafter, the organizational coupling among nearly all services increased from 2018-06-05 to 2019-06-05. However, from 2019-06-05 to 2020-06-05, we can observe the decrease of the coupling amongst all services except \textit{Kayenta} and \textit{Keel}, whose coupling with other services still increased. It implies that there are still developers from other service teams contributing to these two newly established services. From 2020-06-05, we can easily observe the organizational coupling among all services decreases in the last three years. 

\section{Discussion}

In this study, we propose the organizational coupling between microservices as a measure to evaluate how much any two microservices are coupled by the cross-service contribution behaviors of the developers. Such coupling is also damaging to the quality of microservice architecture because spontaneous and unregulated contributions across will inevitably result in an increase in unnecessary communication costs, mismatch between developers and code, and risks in deteriorating system architecture \cite{brooks1995mythical,conway1968committees,fowler202210}. Here we define the organizational coupling of two different microservices as the degree to which the developers cross-contribute between them. The method of evaluating the organizational coupling between two given microservices includes three steps: 1) identifying the contributor team of each microservice and finding the developers who contribute in both; 2) calculating the contribution switch of each common developer and using it as the weight on his/her mean contribution on both microservices; 3) summing all the common developers' weighted cross-service contribution of both microservices as the organizational coupling value. This answers the research question.

When considering the organizational coupling between microservices, we consider that the switching behavior of the developers is a key factor. The reason is that people need to stop thinking about one task in order to fully transition their attention and perform well on another \cite{leroy2009so}. Therefore, the more frequently developers switch between different microservices the more difficult for them to concentrate and perform well on any. Thus, it is reasonable to consider the two microservices organizationally coupled when developers contribute across them as such switching behaviors can influence the quality of both. However, the current calculation is more to take this factor into account as a proof-of-concept rather than accurately calculate the values. So herein, we simply conceptualize the contribution switch as the switch times between two services within a given time without considering the timespan between the switches. Furthermore, we shall also consider other factors when calculating the contribution switch, e.g., microservice priority \cite{bendoly2014prioritizing}, project roles \cite{grotto2022switching}, and so on. 

For future work, we shall continue to enrich the concept of organization coupling by taking into account more factors as parameters. On the other hand, strategies and mechanisms to monitor and handle such organizational couplings are also required in order to continue promoting the principle of ``one microservice per developer". For example, we can adopt time series to monitor the changes in organizational coupling networks together with anomaly detection techniques to identify the severe coupling whenever occurring \cite{shaukat2021review}. Furthermore, we shall also investigate techniques to reduce organizational coupling by encouraging developers to reduce contribution switching frequency or developer number. 

\section{Conclusion}

In this study, we propose the concept of organizational coupling as a measure to evaluate how much any two microservices are coupled by the cross-service contribution behaviors of the developers. Such organizational coupling needs to be detected and mitigated in time to prevent future losses. Therefore, we also propose an automatable approach to evaluate the organizational couple by investigating the microservice ownership and cross-service contribution and validate its usefulness with a case study. Organizational coupling is a critical issue for microservice-based systems on the organizational structural level. Such issues can have a potential impact on the deterioration of system architecture which needs to be detected and addressed in time.

%
%
%

\bibliographystyle{splncs04}
\bibliography{bib}

\end{document}